\documentclass{emulateapj}

\shorttitle{Temperatures in \ion{H}{2} regions}
\shortauthors{Rodr\'\i guez \& Garc\'\i a-Rojas}

\begin{document}

\title{Temperature structure and metallicity in \ion{H}{2} regions}

\author{M\'onica Rodr\'\i guez}
\affil{Instituto Nacional de Astrof\'\i sica, \'Optica y Electr\'onica,
Apdo Postal 51 y 216, 72000 Puebla, Mexico}
\email{mrodri@inaoep.mx}

\author{Jorge Garc\'\i a-Rojas\altaffilmark{1,2}}
\affil{Instituto de Astronom\'\i a, Universidad Nacional Aut\'onoma de M\'exico,
Apdo Postal 70-264, 04510 M\'exico~DF, Mexico}

\altaffiltext{1}{Visiting Astronomer, Instituto Nacional de Astrof\'\i sica,
\'Optica y Electr\'onica, Apdo Postal 51 y 216, 72000 Puebla, Mexico}
\altaffiltext{2}{Current address: Instituto de Astrof\'\i sica de Canarias,
E-38200 La Laguna, Tenerife, Spain}
\email{jogarcia@iac.es}

\begin{abstract}
The metallicities implied by collisionally excited lines (CELs) of heavy elements
in \ion{H}{2} regions are systematically lower than those implied by
recombination lines (RLs) by factors $\sim2$, introducing uncertainties of the
same order in the metallicities inferred for the interstellar medium of any
star-forming galaxy. Most explanations of this
discrepancy are based on the different sensitivities of CELs and RLs to
electron temperature, and invoke either some extra heating mechanism producing
temperature fluctuations in the ionized region or the addition of cold gas in
metal-rich inclusions or ionized by cosmic rays or X rays. These explanations
will change the temperature structure of the ionized gas from the one predicted
by simple photoionization models and, depending on which one is correct,
will imply different metallicities for the emitting gas. We select nine
\ion{H}{2} regions with observed spectra of high quality and show that
simple models with metallicities close to the ones implied by oxygen CELs
reproduce easily their temperature structure, measured with
$T_{\rm e}$([\ion{N}{2}])/$T_{\rm e}$([\ion{O}{3}]), and their oxygen CELs
emission.
We discuss the strong constraints that this agreement places on the possible
explanations of the discrepancy and suggest that the simplest explanation,
namely errors in the line recombination coefficients by factors
$\sim2$, might be the correct one.
In such case, CELs will provide the best estimates of metallicity.
\end{abstract}

\keywords{ISM: abundances --- \ion{H}{2} regions}

\section{Introduction}
\label{int}

In an \ion{H}{2} region, the temperature of a volume of gas in thermal
equilibrium is determined
by the balance between heating, mainly through photoionization, and cooling,
mainly through recombination and radiation of collisionally excited lines (CELs).
The most important coolants are CELs arising from low-energy levels of ions of
abundant metals, like O$^{+}$, O$^{++}$, N$^{+}$, S$^{+}$, and S$^{++}$, and
the gas metallicity is the single most important parameter in the
determination of the final temperature of the gas \citep{ost06}.

Temperatures can be measured using the relative intensities of CELs emitted by
ions like O$^{++}$ and N$^+$, and their associated temperatures,
$T_{\rm e}$([\ion{O}{3}]) and $T_{\rm e}$([\ion{N}{2}]), can be used to
characterize the zones of high and low degree of ionization within a nebula.
These temperatures can be used to calculate line emissivities and hence the
abundances of different elements. The oxygen abundance is probably the most
reliable one that can be derived in \ion{H}{2} regions and it is generally
used as a proxy for the metallicity of the gas.
When optical spectra of high quality are available, the oxygen abundance is
derived by adding the O$^+$/H$^+$ and O$^{++}$/H$^+$ abundance
ratios implied by the intensities of optical [\ion{O}{2}] and [\ion{O}{3}]
lines (relative to hydrogen recombination lines) and the corresponding
emissivities calculated for
$T_{\rm e}$([\ion{N}{2}]) and $T_{\rm e}$([\ion{O}{3}]), respectively.
However, the O$^{++}$ abundance can also be determined using several weak
optical \ion{O}{2} recombination lines (RLs) and these lead to abundances
that are consistently higher than the ones derived from CELs.
A similar result has been found with CELs and RLs of other ions, both in
\ion{H}{2} regions and in planetary nebulae \citep[see, e.g.,][]{est04,liu04},
but with higher uncertainties.
The abundance discrepancy factors (ADFs) for O$^{++}$,
$ADF(\mbox{O}^{++})=(\mbox{O}^{++}/\mbox{H}^{+})_{\rm RL}/
(\mbox{O}^{++}/\mbox{H}^{+})_{\rm CEL}$,
are of the order of 2 for \ion{H}{2}
regions and many planetary nebulae, but can reach much higher values, up to
70, in the latter objects \citep[][and references therein]{liu06}.

Most of the proposed explanations of the ADF are based on the different
dependence with temperature of the emissivities of CELs (strongly increasing
with $T_{\rm e}$) and RLs ($\propto T_{\rm e}^{-m}$ with $m\simeq1$).
If there are large temperature fluctuations in the gas, the emission of CELs
will be heavily weighted towards the hotter regions; the emission of RLs towards
the colder ones.
Depending on the mechanism responsible for the fluctuations, either CELs or RLs
will provide the better estimate of the gas metallicity.
The fluctuations could be due to hot gas heated by shocks, stellar winds, small
dust grains, magnetic reconnection \citep[see][and references therein]{tor03}
or can be due to cold gas in metal-rich inclusions \citep{liu00,tsa05,sta07}
or cold gas ionized by cosmic rays \citep{gia05} or X rays \citep{erc09}.
Some or all of these effects are likely to be present in real objects, but it is
not clear yet whether they can explain the derived ADFs or how much they
contribute to the observed emission.
This introduces an uncertainty in the absolute values of the metallicities in
the interstellar medium.

One way to approach this issue, the one we will follow here, is to build a
series of simple photoionization models with input parameters that can be
considered representative of some observed objects and to see whether the models
reproduce the observed spectra.
We center on nine \ion{H}{2} regions with available spectra of high quality and
use Cloudy \citep{fer98} to calculate grids of photoionization models
that depend only on a few free parameters.
The observed and predicted spectra are analyzed in a similar way to determine 
electron densities, temperatures, and the ionic and total abundances of oxygen
using the intensities of CELs relative to hydrogen lines.
We show that the models reproduce easily the general trends defined by the
observations.
We argue that they are representative of the observed objects and discuss how
far they can be modified by adding some extra ingredient that could also explain
the intensities of \ion{O}{2} RLs without destroying the agreement.\\

\section{The observational sample}
\label{obs}

We consider nine \ion{H}{2} regions with published spectra of the best
quality: the Galactic \ion{H}{2} regions M42 \citep{est04}, NGC~3576
\citep{gar04}, NGC~2467 \citep{gar05}, M16, M20, NGC~3603 \citep{gar06},
M8 and M17 \citep{gar07}, and 30~Doradus \citep{pei03} in the Large Magellanic
Cloud (LMC).
Deep spectra of these nebulae were obtained with the VLT UVES echelle
spectrograph covering the range 3100--10400 \AA\ with spectral resolution of
$\Delta\lambda\simeq8800/\lambda$.
The spectra, which were extracted over small fractional areas,
$3''\times8\farcs5$ or $3''\times10''$, allowed the measurement of 235--555
emission lines in each nebula, including several weak \ion{O}{2} ORLs, that were
detected with good signal-to-noise ratios.
Physical conditions could be derived from several diagnostics that led to
broadly consistent values.

Since we will be comparing the values of the temperatures and abundances derived
from the observed spectra with those derived from the line intensities predicted
by the photoionization models, we decided to use the same atomic data for all
calculations.
Hence, we recomputed the parameters of interest with the {\sl nebular} package
in IRAF\footnote{IRAF is distributed by NOAO, which is operated by AURA, Inc.,
under cooperative agreement with NSF.} but using the same atomic data
as Cloudy.
This implied changing the tables of transition probabilities and
collision strengths used in IRAF for the elements of interest (references for the
values we used for these atomic data can be found in \citealt{gar09}).
We did not change the number of levels used in the calculations nor the
procedure used by IRAF to get the emissivities of the hydrogen lines, but the
effects of these differences are small.

The recomputed electron densities, $n_e$, are a weighted mean of the values
implied by the diagnostics [\ion{S}{2}] $\lambda6716/\lambda6731$, [\ion{O}{2}]
$\lambda3726/\lambda3729$, and [\ion{Cl}{3}] $\lambda5517/\lambda5537$.
The values of $T_e$([\ion{O}{3}]) are based on the ratio of line intensities
$(\lambda4959+\lambda5007)/\lambda4363$; those of $T_e$([\ion{N}{2}]) on
$(\lambda6548+\lambda6583)/\lambda5755$.
In some of the previous studies, the values of $T_e$([\ion{N}{2}]) were
corrected for the contribution of recombination to the intensity of
[\ion{N}{2}]~$\lambda5755$ \citep{rub86}, but since the corrections
are based on uncertain estimates of the abundance of N$^{++}$, we prefer to
apply that correction to the models (see Section~\ref{mod}).
The values of $I(\mbox{[\ion{O}{2}]~}\lambda\mbox{3726+29})/I(\mbox{H}\beta)$,
$n_e$, and $T_e$([\ion{N}{2}]) are used to derive the
$\mbox{O}^+/\mbox{H}^+$ abundance ratio;
$I(\mbox{[\ion{O}{3}]~}\lambda4959+\lambda5007)/I(\mbox{H}\beta)$,
$n_e$, and $T_{\rm e}$([\ion{O}{3}]) are
used for $\mbox{O}^{++}/\mbox{H}^+$.
The total abundance of oxygen is then obtained from the sum of the O$^+$ and
O$^{++}$ ionic abundances.
We also recomputed the O$^{++}$ abundances implied by RLs using the \ion{O}{2}
multiplet M1 and the recombination coefficients of
\citet{peq91} which are those implemented in Cloudy.
In Table~\ref{t1} we show the physical conditions, $T_e$, $n_e$ and the
degree of ionization, the O/H abundances derived from CELs and
the ADFs for all the sample objects.
Despite the slight changes in the procedure and in the atomic data used, the
final values of the oxygen abundance and of the ADF are consistent within the
errors with the values derived in the previous analyses.\\

\begin{deluxetable*}{llllccc}
\tabletypesize{\footnotesize}
\tablecolumns{7}
\tablecaption{Physical conditions and oxygen abundances for the sample objects
 \label{t1}}
\tablewidth{0pt}
\tablehead{
\colhead{ } & \colhead{$n_e$} &
    \colhead{$T_e$([\ion{N}{2}])} & \colhead{$T_e$([\ion{O}{3}])} &
    \colhead{ } & \colhead{ } & \colhead{ } \\
\colhead{Object} & \colhead{(cm$^{-3}$)} & \colhead{(K)} &
    \colhead{(K)} & \colhead{$12+\log$(O/H)} &
    \colhead{$\log(\mbox{O}^+/\mbox{O}^{++})$} & \colhead{$ADF(\mbox{O}^{++})$}
}
\startdata
\objectname{M8}     & $1550\pm150$     & $8500\pm120$      & $8020\pm100$
&$8.44\pm0.03$  & $+0.40\pm0.05$ &$2.2\pm0.2$ \\
\objectname{M16}     & $980\pm120$     & $8500\pm150$      & $7580\pm150$
&$8.52\pm0.03$  & $+0.45\pm0.06$ &$2.4\pm0.6$ \\
\objectname{M17}     & $420\pm80$    & $9150\pm250$      & $7950\pm100$
&$8.53\pm0.02$  & $-0.76\pm0.06$ &$1.8\pm0.2$ \\
\objectname{M20}     & $220\pm40$    & $8500\pm150$      & $7750\pm250$
&$8.48\pm0.03$  & $+0.64\pm0.07$ &$1.9\pm1.2$ \\
\objectname{M42}     & $6800\pm600$     & $10100\pm250$     & $8250\pm60$
&$8.53\pm0.02$  & $-0.68\pm0.05$ &$1.50\pm0.05$ \\
\objectname{NGC~2467}     & $260\pm50$    & $9600\pm200$        & $8900\pm100$
&$8.35\pm0.03$  & $+0.35\pm0.05$ &$1.8\pm0.2$ \\
\objectname{NGC~3576}     & $1400\pm200$     & $8950\pm200$      &
$8400\pm80$   &$8.53\pm0.02$  & $-0.40\pm0.06$ &$2.0\pm0.3$ \\
\objectname{NGC~3603}     & $2350\pm400$     & $11650\pm570$     &
$9000\pm150$  &$8.47\pm0.03$  & $-1.21\pm0.09$ &$1.9\pm0.3$ \\
\objectname{30~Doradus} & $380\pm50$    & $10800\pm300$        &
$9850\pm100$  &$8.36\pm0.02$  & $-0.77\pm0.06$ &$1.5\pm0.1$ 
\enddata
\end{deluxetable*}

\section{The photoionization models}
\label{mod}

Calculations were performed with version 08.00 of Cloudy, last described
by \citet{fer98}.
Taking into account that the observations covered small fractional areas of the
objects, that different ions lead to similar densities, and that the line
emissivities are roughly proportional to the square of the density,
one-dimensional plane-parallel models with constant density should be a good
approximation.

We use models with total hydrogen densities close to the electron densities
derived for the observed objects: $n_{\rm H}=300$ cm$^{-3}$ for
M17, M20, 30~Doradus, and NGC~2467; $n_{\rm H}=1000$ cm$^{-3}$ for NGC~3576, M8,
and M16; and $n_{\rm H}=5000$ cm$^{-3}$ for NGC~3603 and M42.

We define the ionizing radiation field through two parameters: the effective
temperature of the ionizing star, $T_{\rm eff}$, and the ionization parameter
$u=\phi_{\rm H}/n_{\rm H}c$, with $\phi_{\rm H}$ the number of hydrogen
ionizing photons arriving at the inner face of the cloud per unit area and unit
time, and $c$ the speed of light.

We consider the two sets of state-of-the-art model atmospheres for O-type stars
included in Cloudy: {\sc wm-basic} \citep{pau01} and {\sc tlusty}
\citep{lan03}.
Comparisons between the ionizing fluxes predicted by different stellar models
and discussions on the effects on the resulting photoionization models can be
found in \citet{mar05}, \citet{sim08}, or in the references therein.
Here we present the results obtained with {\sc wm-basic} and use the results
obtained with {\sc tlusty} as a check for robustness.
We use stellar models with surface gravity $\log(g)=4.0$, solar metallicity,
and effective temperatures $T_{\rm eff}=35000$, $37000$,
$39000$, $41000$, $43000$, $45000$, and $50000$~K.
The stellar models with $T_{\rm eff}=35000$~K to 45000~K represent roughly the
spectral types from O3 to O8 \citep{mar05}; the models with
$T_{\rm  eff}=50000$~K probe the effects of more massive stars.
The values of $u$ are chosen for each value of $T_{\rm eff}$ so that the set of
models cover the range of degrees of ionization found for the observed objects.
For the models with $n_{\rm H}=300$ and 1000 cm$^{-3}$ we use
$\log(u)=-1.5$, $-1.0$, $-0.5$, and $0.0$ for $T_{\rm eff}=35000$ K;
$\log(u)=-2.5$, $-2.0$, $-1.5$, and $-1.0$ for $T_{\rm eff}=37000$ K;
$\log(u)=-3.0$, $-2.5$, $-2.0$, and $-1.5$ for $T_{\rm eff}=39000$ K; and
$\log(u)=-3.5$, $-3.0$, $-2.5$ and $-2.0$ for $T_{\rm eff}=41000$, $43000$,
$45000$, and $50000$~K.
For the models with $n_{\rm H}=5000$ cm$^{-3}$, we increased the values
of $\log(u)$ by 0.5~dex.

We also computed models ionized by two stars with $T_{\rm eff}=35000$ and
$45000$~K that provide equal amounts of hydrogen ionizing photons
($\log(u)=-3.3$, $-2.8$, $-2.3$, or $-1.8$ for each star). These models
illustrate the effect of a composite stellar radiation field.

We use the set of abundances labeled as ``\ion{H}{2} region'' (or ``Orion'')
in Cloudy and the associated ``\ion{H}{2} region'' dust grains.
We scaled both metals and grains in the ``\ion{H}{2} region'' set by the same
factor
in order to
explore different metallicities. We changed the metallicity using steps of
0.1~dex and explored the range $12+\log(\mbox{O}/\mbox{H})=8.3\mbox{--}9.1$.

By default, Cloudy stops when the model reaches an electron temperature of
4000~K. However, since some of the models with high metallicity reach this
temperature in the fully ionized zone, we switched off this criterion and
used instead the condition $\log(n_{\rm e}/n_{\rm H})\le-0.5$.

For each model we get a list with the intensities relative to H$\beta$ of the
lines of interest.
We introduce this list in IRAF and use it to calculate physical conditions and
ionic abundances following the same procedure we used in the analysis of the
observed objects (see Section~\ref{obs}).

As mentioned in Section~\ref{obs}, the upper level of the
[\ion{N}{2}]~$\lambda5755$ line used to derive $T_{\rm e}$([\ion{N}{2}]) can be
populated by recombination processes \citep{rub86}.
If this effect is not taken into account, the values of
$T_{\rm e}$([\ion{N}{2}]) derived for the models will be lower than those derived
for the observed objects, especially for the cases where the degree of ionization
is high.
\citet{liu00} estimated the contribution to the intensity of this line in terms
of the N$^{++}$ abundance. We used their correction to add this contribution to
the intensity of the [\ion{N}{2}] in each model using the N$^{++}$ and H$^+$
column densities and $T_{\rm e}=10000$ K
(the dependence on $T_{\rm e}$ is very weak).
Then we calculated a second set of physical conditions and abundances for all
the models. Since the correction is just an estimate, both the corrected and
uncorrected results are shown below.\\

\section{Results}
\label{res}

We compared the physical conditions and abundances derived for the models with
those derived for the observed objects.
Most of the models lead to derived abundances, (O/H)$_{\rm out}$, that are lower
than the input ones.
This is a well known effect (\citealt{sta05} and references therein),
due to the exponential dependence with temperature of the CELs used to derive
$T_{\rm e}$ and the O$^+$ and O$^{++}$ abundances.
In the presence of a temperature gradient, this dependence leads to an
overestimation of $T_{\rm e}$ and hence of the emissivity of the CELs. This
translates into an underestimation of the derived abundances relative to
hydrogen.
As discussed by \citet{sta05}, under some circumstances, models with high
metallicity and low density develop a gradient so steep that the difference
between the input (O/H) and (O/H)$_{\rm out}$ becomes substantial.
One extreme case is our model with $n_{\rm H}=300$ cm$^{-3}$,
$T_{\rm eff}=41000$~K, $\log(u)=-2.0$, and $12+\log(\mbox{O}/\mbox{H})=9.1$.
The abundance derived from the analysis of the spectrum predicted for this model
is $12+\log(\mbox{O}/\mbox{H})_{\rm out}=8.55$, and $ADF(\mbox{O}^{++})\simeq4$.
Other high-metallicity models have ADFs close to the ones derived for the
observed objects, but also $T_{\rm e}(\mbox{[\ion{O}{3}]})\la6000$ K, much
lower than the observed values.
All the models that consistently give values of (O/H)$_{\rm out}$,
$T_{\rm e}$([\ion{N}{2}]), and $T_{\rm e}$([\ion{O}{3}]) close to the observed
ones, have input metallicities 0 to 0.06~dex higher than the derived ones and
$ADF(\mbox{O}^{++})\simeq1.1$.

\begin{figure*}
\plotone{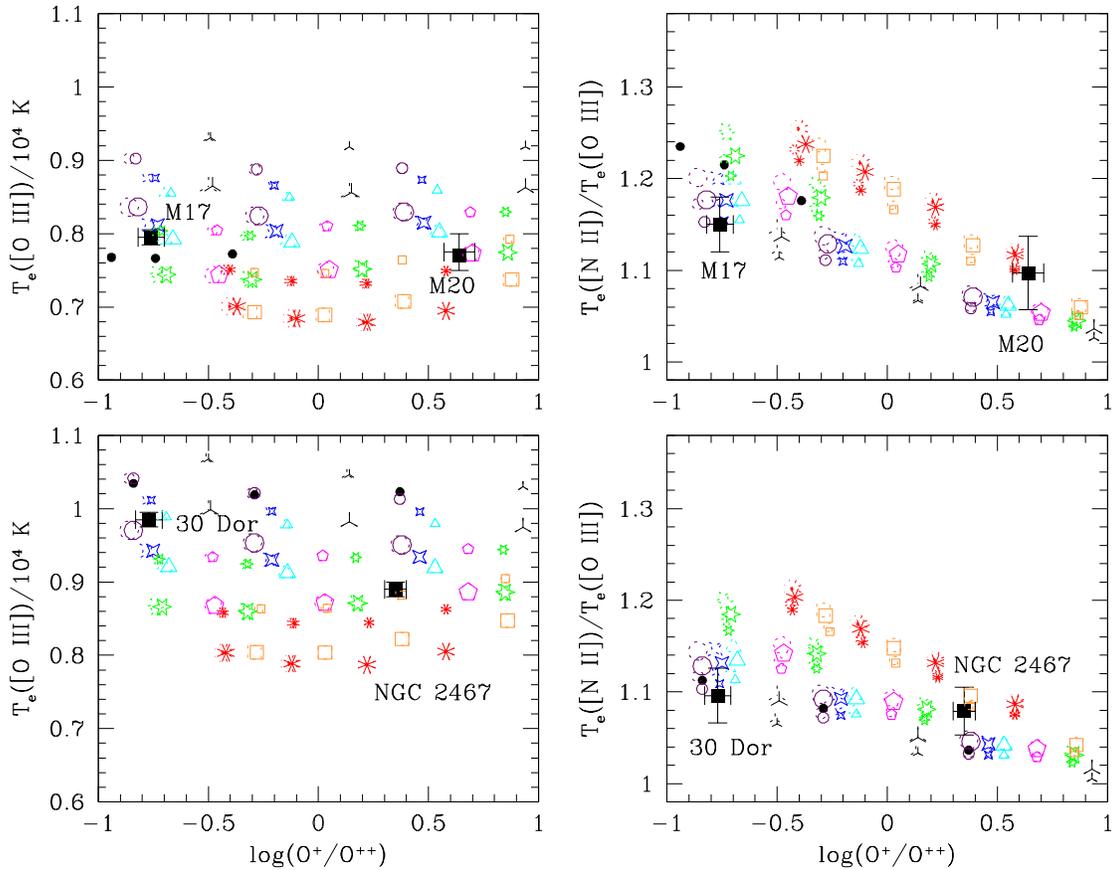}
\caption{Values of $T_{\rm e}$([\ion{O}{3}]) and
$T_{\rm e}$([\ion{N}{2}])/$T_{\rm e}$([\ion{O}{3}]) for observed objects (filled
black squares) and photoionization models (other symbols). The models have
$n_{\rm H}=300$ cm$^{-3}$ and $T_{\rm eff}=35000$~K (red asterisks),
$37000$~K (orange open squares), $39000$~K (green six-pointed stars),
$41000$~K (cyan triangles), $43000$~K (blue four-pointed stars),
$45000$~K (violet circles), $50000$~K (black inverted Ys),
or both $T_{\rm eff}=35000$~K and $45000$~K
(magenta pentagons). Big symbols correspond to models with
$12+\log(\mbox{O}/\mbox{H})=8.6$ (top panels) and 8.4 (bottom panels). Small
symbols have $12+\log(\mbox{O}/\mbox{H})=8.5$ (top panels) and 8.3 (bottom
panels). Dotted symbols show values derived considering the contribution of
recombination to the intensity of [\ion{N}{2}]~$\lambda5755$.
The small filled circles show the effects of intervening material changing the
shape of the ionizing radiation field (top panels, for models with
$T_{\rm eff}=39000$~K and $12+\log(\mbox{O}/\mbox{H})=8.6$) and of decreasing
the abundances of C, N, S, and Ne relative to O (bottom panels, for models with
$T_{\rm eff}=45000$~K and $12+\log(\mbox{O}/\mbox{H})=8.4$).
See the text for more explanations.\label{fig1}\\}
\end{figure*}

\begin{figure*}
\plotone{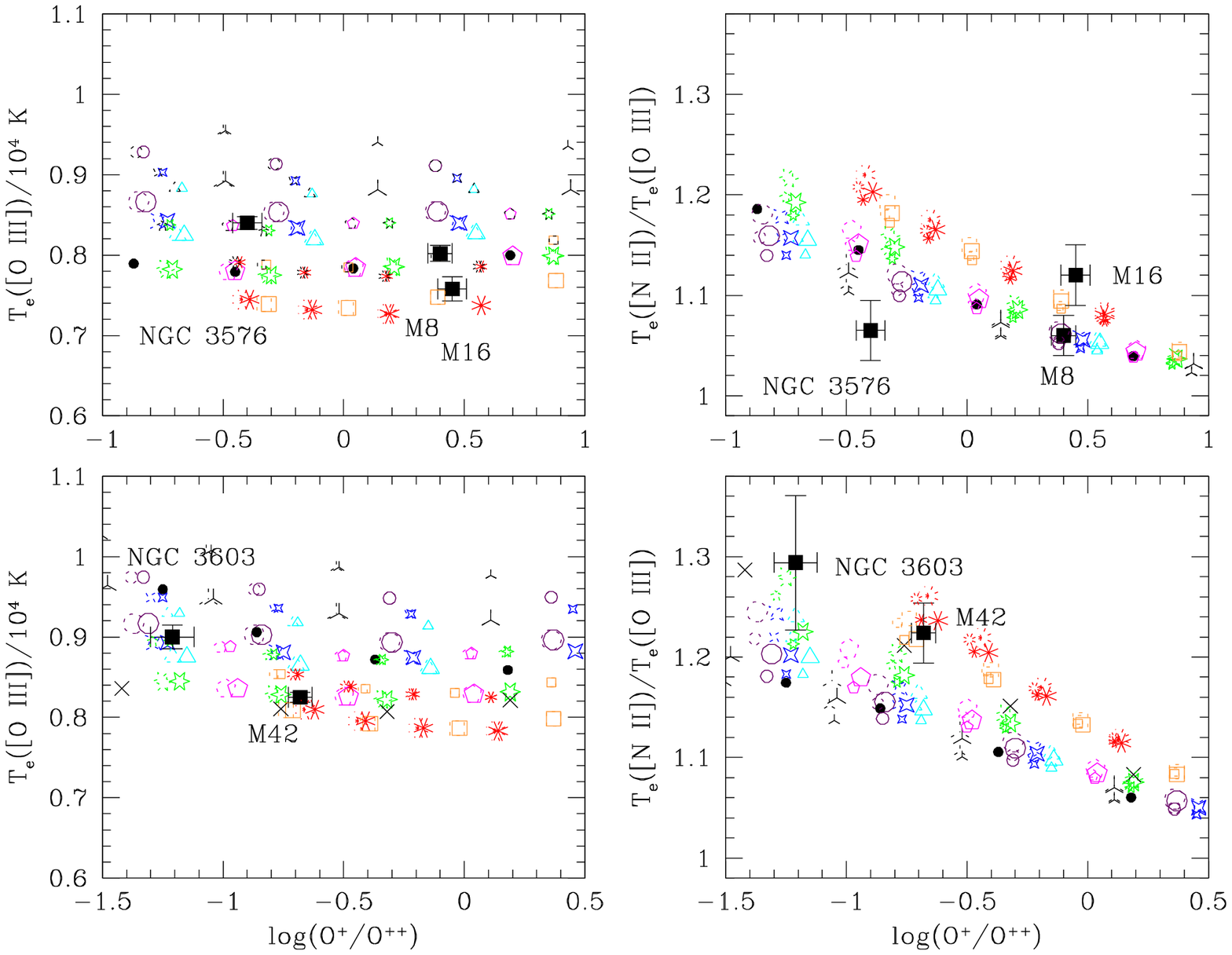}
\caption{Values of $T_{\rm e}$([\ion{O}{3}]) and
$T_{\rm e}$([\ion{N}{2}])/$T_{\rm e}$([\ion{O}{3}]) for observed objects (filled
black squares) and photoionization models (other symbols). The models have
$n_{\rm H}=1000$ cm$^{-3}$ (top panels) or 5000 cm$^{-3}$ (bottom
panels), and $T_{\rm eff}=35000$~K (red asterisks),
$37000$~K (orange open squares), $39000$~K (green six-pointed stars),
$41000$~K (cyan triangles), $43000$~K (blue four-pointed stars),
$45000$~K (violet circles), $50000$~K (black inverted Ys),
or both $T_{\rm eff}=35000$~K and $45000$~K (magenta pentagons).
Big symbols correspond to models with $12+\log(\mbox{O}/\mbox{H})=8.6$; small
symbols have $12+\log(\mbox{O}/\mbox{H})=8.5$.
Dotted symbols show values derived considering the contribution of
recombination to the intensity of [\ion{N}{2}]~$\lambda5755$.
The small filled circles show the effects of decreasing the metallicity of the
ionizing stars by a factor of 2 (top panels) and of changing the dust grains
from ``Orion'' type to ``ISM'' type (bottom panels) for models with
$T_{\rm eff}=39000$~K and $12+\log(\mbox{O}/\mbox{H})=8.6$. Models with
$T_{\rm eff}=39000$~K and $12+\log(\mbox{O}/\mbox{H})=8.6$ but with no grains
included are shown as crosses in the bottom panels.
See the text for more explanations.\label{fig2}\\}
\end{figure*}

Figures~\ref{fig1} and \ref{fig2} show a comparison of $T_{\rm e}$([\ion{O}{3}])
and $T_{\rm e}$([\ion{N}{2}])/$T_{\rm e}$([\ion{O}{3}]) as a function of the
derived values for the degree of ionization, $\log(\mbox{O}^{+}/\mbox{O}^{++})$,
for models and observations. The results implied by models with two input
metallicities are shown for each object:
$12+\log(\mbox{O}/\mbox{H})=8.5$ and $8.6$ for M8, M16, M17, M20, NGC~3576, M42,
and NGC~3603; $12+\log(\mbox{O}/\mbox{H})=8.3$ and $8.4$ for NGC~2467 and
30~Doradus. The output metallicities of these models, (O/H)$_{\rm out}$, bracket
the observed values listed in Table~\ref{t1}. 
The observed objects are compared with those models that have similar values of
$n_{\rm e}$.
It can be seen that these models also reproduce easily the temperatures derived
for the observed nebulae.
An exception is the value of $T_{\rm e}$([\ion{N}{2}])/$T_{\rm e}$([\ion{O}{3}])
for NGC~3576.
This could be just a statistical fluctuation, with one in nine measurements
deviating with a $2\sigma$ error bar from the expected result, or could indicate
that something is different in the environment of this object.

Note that the models ionized by two stars with $T_{\rm eff}=35000$~K and
$45000$~K (the pentagons in Figures~\ref{fig1} and \ref{fig2}) give results that
are intermediate between those shown by the models ionized by one star with
either $T_{\rm eff}=35000$~K or $45000$~K (the asterisks and the circles,
respectively).
Hence, models ionized by composite spectra, with several values of $T_{\rm eff}$,
will also appear in the regions sampled by the single $T_{\rm eff}$ models
in Figures~\ref{fig1} and \ref{fig2}.

Very similar results to the ones found with {\sc wm-basic} were obtained using
the {\sc tlusty} model atmospheres.
Although individual models would appear in somewhat different positions if
plotted in Figures~\ref{fig1} and \ref{fig2}, the areas covered by both kinds of
models are roughly similar.

Many simplifying assumptions were made constructing these models.
The agreement with the observations suggests that despite the simplifications
and uncertainties, the models have captured some basic characteristics of the
observed objects.
We have already addressed the effects of having a composite ionizing spectra
and of using different model atmospheres.
We now address other uncertainties and simplifications of the models to further
explore how general are the results.

\subsection{Effect of intervening material}

Our models assume that there is no intervening material between the ionizing
stars and the observed area. However, this material, if present, will change
the shape (and luminosity) of the ionizing radiation field reaching the
observed region and, in principle, this could lead to departures from the
behavior of the models in Figures~\ref{fig1} and \ref{fig2}.
To check for this effect we calculated several models where the stellar
radiation field was first passed through a slab of material of fixed depth and
the transmitted radiation was then used to ionize the slab of interest.

We start considering intervening material with $n_{\rm H}=300$ cm$^{-3}$,
$12+\log(\mbox{O}/\mbox{H})=8.6$, and a radiation field with
$T_{\rm eff}=39000$~K and $\log(u)=-1.0$.
If enough material is present, these conditions will produce a depth
of ionized material of about 4 pc.
We consider slabs with thicknesses of 1, 2, and 3 pc and use the radiation
transmitted through them to ionize material with the same $n_{\rm H}$ and
metallicity.
These models have values of the ionization parameter of $\log(u)=-1.23$, $-1.5$,
and $-1.93$, respectively.
The spectra produced by the latter slabs were analyzed in the same manner as the
original models and the results are plotted as small filled circles in
the top panels of Figure~\ref{fig1}. It can be seen that they give results
similar to the original grid of models.

We calculated similar models (with $12+\log(\mbox{O}/\mbox{H})=8.6$,
$T_{\rm eff}=39000$~K, and initial $\log(u)=-1.0$) for material with
$n_{\rm H}=1000$ cm$^{-3}$. The incident radiation field was first passed
through slabs with the same density and depths of 0.3, 0.6, and 0.9 pc (so that
for the final models $\log(u)=-1.23$, $-1.5$, and $-1.9$).
We did other tests with intervening material of lower density, $n_{\rm H}=100$
cm$^{-3}$, and depths of 1, 2, and 3 pc (leading to $\log(u)=-1.06$, $-1.11$,
and $-1.17$ for the final models with $n_{\rm H}=1000$ cm$^{-3}$).
All these models also gave similar results to
those produced by the original grid of models.

\subsection{Changing the relative abundances of some elements}

We are using the set of ``\ion{H}{2} region'' chemical abundances in Cloudy
\citep[based on the abundances derived for M42 by][]{bal91,rub91,ost92}
and scaling all of them by the same amount to change the overall metallicity
of the gas, which we measure through O/H.
For the other elements, not all ionization states show lines in the optical and
their total abundances are derived using ionization correction factors that make
them less reliable.
Changing the abundances relative to oxygen of those elements
that contribute significantly to the cooling might change our results.
We compared the values of the C/O, N/O, S/O, and Ne/O abundance ratios in
Cloudy with the values derived for these ratios in the original papers
where the analysis of the observed objects was first presented.
The most discrepant results are found for 30~Doradus, with C/O, N/O, S/O, and
Ne/O lower than the Cloudy values by 0.4, 0.5, 0.1, and 0.14 dex,
respectively.
Hence, we ran three new models with $n_{\rm H}=300$ cm$^{-3}$,
$12+\log(\mbox{O}/\mbox{H})=8.4$, $T_{\rm eff}=45000$~K, $\log(u)=-3.0$
$-2.5$, and $-2.0$, and with the abundances of C, N, S, and Ne scaled down by
the amounts listed above.
The results are shown as small filled circles in the bottom panels of
Figure~\ref{fig1}.
As expected, they show somewhat higher values of $T_{\rm e}$, but the
values of $T_{\rm e}$([\ion{N}{2}])/$T_{\rm e}$([\ion{O}{3}]) do not change
significantly.

\subsection{Changing the metallicity of the ionizing star}

All the calculations were performed with stellar spectra for solar
metallicity, with $12+\log(\mbox{O}/\mbox{H})_{\odot}\ga8.7$
\citep[e.g.][and references therein]{sco09}.
However, the models have input metallicities in the range
$12+\log(\mbox{O}/\mbox{H})_{\odot}=8.3\mbox{--}8.6$.
We calculated four new models with $n_{\rm H}=1000$ cm$^{-3}$ and
$12+\log(\mbox{O}/\mbox{H})=8.6$, ionized by stars with $T_{\rm eff}=39000$~K
and half-solar metallicity, and with $\log(u)=-3.0$ $-2.5$, $-2.0$, and $-1.5$.
The results of these models are plotted as small filled circles in the top
panels of Figure~\ref{fig2}.
The new models have a higher degree of ionization but show a similar behavior to
the original ones.

\subsection{Effect of dust grains}

The models include dust grains, and the grains affect the temperature, mainly
through heating by the photoelectric effect.
The fractional contribution of grains to the heating increases with the
ionization parameter and decreases with the effective temperature of the
ionizing star.
For the main photoionization models described so far, the contribution can be
negligible or reach $\sim30$\% of the global heating.
We are using the ``\ion{H}{2} region'' (or ``Orion'') dust grains, a mixture of
graphite and silicate grains that is deficient in small particles in order to
explain the grey extinction in M42 \citep{bal91}, but
there are many uncertainties related to which grains would be suitable.
Changing the type of dust grains will change the temperature structure.
To illustrate this effect, we calculated four new models with
$n_{\rm H}=5000$ cm$^{-3}$, $12+\log(\mbox{O}/\mbox{H})=8.6$,
$T_{\rm eff}=39000$~K, and $\log(u)=-2.5$, $-2.0$, $-1.5$, and $-1.0$,
changing the type of dust from ``\ion{H}{2} region'' to ``ISM'' (interstellar
medium).
The ``ISM'' dust grains are also composed of graphite and silicate, but
have a size distribution that includes small grains.
The results of these models are plotted as small filled circles in the bottom
panels of Figure~\ref{fig2}.
They have higher values of $T_{\rm e}$([\ion{O}{3}]) and lower
values of $T_{\rm e}$([\ion{N}{2}])/$T_{\rm e}$([\ion{O}{3}]).
Whereas the original models (with $T_{\rm eff}=39000$~K,
$12+\log(\mbox{O}/\mbox{H})=8.6$, and ``\ion{H}{2} region'' dust) had 8\% as
the maximum contribution of grains to the global heating, the new models reach
30\%.
Note however, that the results are not very different from the original ones.

The same test models were computed with no grains included (but keeping the
depleted abundances of refractory elements).
These models show lower values for the derived electron temperatures and higher
values of $T_{\rm e}$([\ion{N}{2}])/$T_{\rm e}$([\ion{O}{3}]) than the original
ones. This is shown by the crosses in the bottom panels of Figure~\ref{fig2}.

One of our objects, 30~Doradus, is located in the LMC, where dust grains can be
very different from Galactic grains. However, the amount of dust in 30~Doradus
is probably well reproduced by the comparison models of the bottom
panels in Figure~\ref{fig1}. All our models have high depletion factors of
refractory elements onto dust grains, and the depletion factors are kept
constant by scaling together the abundances of metals and grains. If we consider
iron a good representative of refractory elements, this assumption is supported
by the high iron depletion factors found in 30~Doradus and several Galactic
\ion{H}{2} regions \citep{rod05}. On the other hand, the grains in 30~Doradus
could have a different composition and be smaller than their Galactic
counterparts \citep[see][for some recent results]{slo08,par09}, but the effects
of the different characteristics of grains in 30~Doradus on the calculated
temperatures are likely to be of the order of those discussed above. For a more
thorough investigation of the effects of changes in composition and size
distribution on model output see \citet{vanh04}.

\subsection{Advection}

The models we are calculating are static.
In the dynamical case, the ionization of neutral material entering the
ionization front (advection), can increase the temperature of the gas in that
region \citep[e.g.][]{rod98}.
This will translate into higher values of $T_{\rm e}$([\ion{N}{2}]),
but the effect is probably small, as discussed by \citet{hen05}.\\

\section{Results for the case with extra heating}

We have shown that the grids of simple photoionization models reproduce very
well the temperatures and metallicities implied by the intensities of CELs
relative to hydrogen recombination lines in the observed objects.
However, these models have $ADF(\mbox{O}^{++})\simeq1.1$ and hence do not
reproduce the intensities of \ion{O}{2} RLs.
In order to reproduce the intensities of both CELs and RLs, some extra ingredient
must be added to the models.
What characteristics should this extra ingredient have so that it does not
destroy the agreement found so far between observations and models?

We start exploring the case where the extra ingredient is a heating agent that
acts in some regions of the nebula producing temperature fluctuations.
In this case, the emission of CELs will be strongly biased towards the hotter
regions leading to overestimates of the derived temperatures and underestimates
of the abundances.
The abundances derived using \ion{O}{2} RLs and the temperature fluctuations
formalism first developed by \citet{pei67} \citep[see also][and references
therein]{pei04} will be more reliable.
As an example, in M8 and M16 the oxygen abundances derived with this procedure
are $12+\log(\mbox{O}/\mbox{H})\simeq8.7\mbox{--}8.8$ \citep{gar06,gar07}.

We modified the function that can be used to introduce a term of extra heating in
Cloudy so that the term was only added at localized depths within the
ionized slab. In this way we can produce spikes in the temperature
structure of the ionized region: the spikes appear in those zones where the term
of extra heating is different from zero.
The approach is somewhat similar to the one used by \citet{bin01}.
The temperature fluctuations due to these spikes introduce a bias in the
temperatures and abundances derived from CELs that produce in turn values of
$ADF(\mbox{O}^{++})$ significantly larger than 1. Note that as long as the spikes
are regularly distributed across the ionized region, models with many thin spikes
lead to results similar to those obtained from models where these spikes are
grouped producing fewer, but broader, spikes.
If one fixes the total number of spikes appearing in the ionized slab, the
required values for the ADF can be obtained by varying the
amount of extra heating through changes in the width or the height of the spikes.

\begin{figure}
\plotone{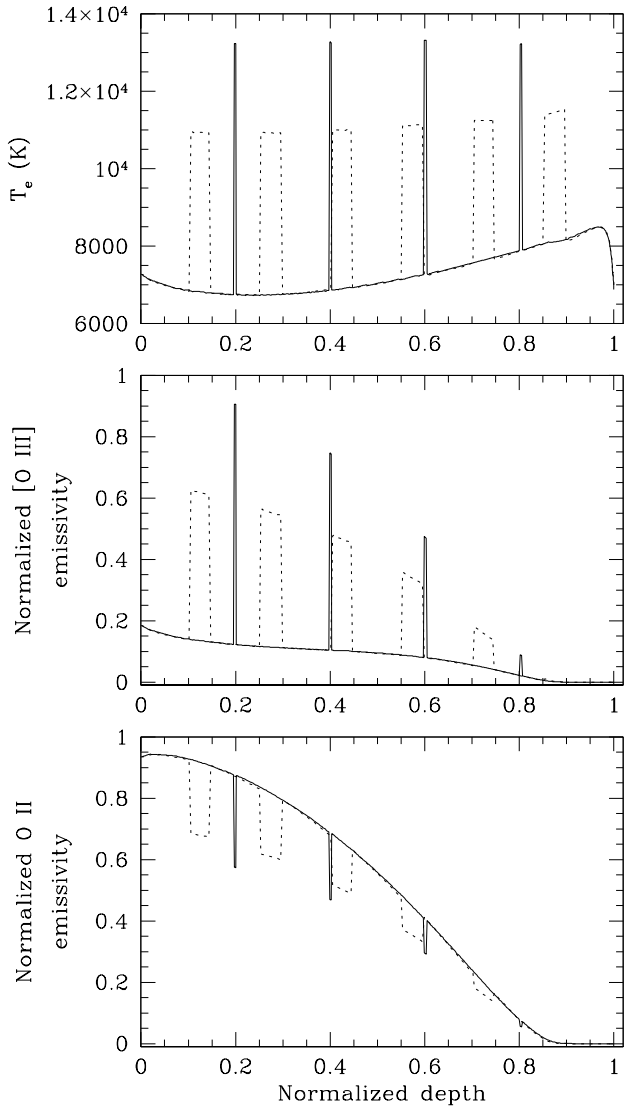}
\caption{Upper panel: Values of $T_{\rm e}$ for two models with terms of extra
heating added in selected regions, as a function of normalized depth into the
ionized slab (with depths of $2.49\times10^{17}$~cm for the model plotted with
a solid line and $2.66\times10^{17}$~cm for the model plotted with a dotted
line). Both models have $T_{\rm eff}=39000$~K, $\log(u)=-2.5$, $n_{\rm H}=1000$
cm$^{-3}$, and input $12+\log(\mbox{O}/\mbox{H})=8.75$. Both models produce
$ADF(\mbox{O}^{++})\sim2$. Middle and lower panels: Emissivities of
[\ion{O}{3}]~$\lambda5007$ and multiplet 1 of \ion{O}{2} for the same models in
arbitrary units.\label{fig3}\\}
\end{figure}

The upper panel of Figure~\ref{fig3} shows the temperature structure of two
models that produce $ADF(\mbox{O}^{++})\simeq2$
for some of the conditions suitable to describe M8 or M16.
The middle and lower panels of Figure~\ref{fig3} show for these two models
the emissivities of [\ion{O}{3}]~$\lambda5007$ and multiplet 1 of \ion{O}{2}
with arbitrary normalization.
Since the models have almost constant electron densities, the emissivities are
modulated by the ionization fraction of O$^{++}$ and by the electron
temperature.
These models illustrate how large the temperature fluctuations must be in order
to produce the required ADFs.
The models have $T_{\rm eff}=39000$~K, $\log(u)=-2.5$,
$n_{\rm H}=1000$ cm$^{-3}$, and input $12+\log(\mbox{O}/\mbox{H})=8.75$.
Derived values are: $\log(\mbox{O}^{+}/\mbox{O}^{++})\simeq0.4$,
$12+\log(\mbox{O}/\mbox{H})=8.56$ and 8.60 (for the models with the broad and
thin spikes, respectively), and $T_{\rm e}$([\ion{N}{2}]$)=9300$~K,
$T_{\rm e}$([\ion{O}{3}]$)=9800$~K (broad spikes),
$T_{\rm e}$([\ion{N}{2}]$)=8000$~K, $T_{\rm e}$([\ion{O}{3}]$)=8200$~K
(thin spikes).
In the model with broad spikes, the extra heating contributes 32\% of the
model's total heating; in the model with thin spikes this contribution is just
6\%.

Models with a term of extra heating just 20\% lower than the one used for the
model with broad spikes in Figure~\ref{fig3} do not have enough temperature
fluctuations to reach $ADF(\mbox{O}^{++})\sim2$: they have
$ADF(\mbox{O}^{++})\la1.8$ for spikes of any width.
On the other hand, we did not explore models with terms of extra heating larger
than the one used for the model with thin spikes in Figure~\ref{fig3}. Those
models would require very short integration steps in Cloudy. Besides, the thin
spikes would dissipate very quickly, as discussed below.

\begin{figure}
\plotone{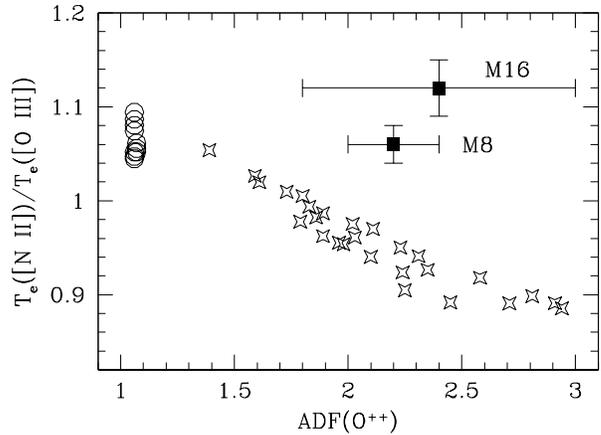}
\caption{Values of $T_{\rm e}$([\ion{N}{2}])/$T_{\rm e}$([\ion{O}{3}]) as a
function of ADF(O$^{++}$) for models with $T_{\rm eff}=39000$~K,
$\log(u)=-2.5$, $n_{\rm H}=1000$ cm$^{-3}$, input
$12+\log(\mbox{O}/\mbox{H})=8.75$, and temperature spikes of different
thickness and widths (four-pointed stars). Also plotted are those models from
the top panels in Figure~\ref{fig2} that have
$\log(\mbox{O}^{+}/\mbox{O}^{++})\sim0.45$ (open circles), a value similar to
those shown by M8 and M16 (plotted with filled squares).\label{fig4}\\}
\end{figure}

The result found for the two models in Figure~\ref{fig3},
$T_{\rm e}$([\ion{N}{2}]$)<T_{\rm e}$([\ion{O}{3}]),
is characteristic of all the models that can be constructed in this way to
produce large ADFs. This can be seen in Figure~\ref{fig4}, where we plot the
values of $T_{\rm e}$([\ion{N}{2}])/$T_{\rm e}$([\ion{O}{3}]) as a
function of ADF(O$^{++}$) for several models with temperature spikes of different
thicknesses and widths. Also shown are the values derived for the original models
with no extra heating and plotted in the top panels of Figure~\ref{fig2} with
$\log(\mbox{O}^{+}/\mbox{O}^{++})\sim0.45$ (a similar degree of ionization to the
ones found in M8 and M16).
None of the models with extra heating can reproduce the observed values of the
temperature ratio, both because the extra heating tends to make similar the
average temperatures of the [\ion{N}{2}] and [\ion{O}{3}] emitting regions, and
because the [\ion{O}{3}] diagnostic ratio is more sensitive to increments in
temperature than the [\ion{N}{2}] ratio.
In order to get $T_{\rm e}$([\ion{N}{2}]$)>T_{\rm e}$([\ion{O}{3}]) as in the
observed objects, the term of extra heating would have to increase with depth
into the cloud.
This explanation of the ADF would also work if the temperature spikes are created
continuously across the full ionized region but dissipate more easily in the
regions closer to the star.
Any attempt to explain the abundance discrepancy with temperature fluctuations
due to extra heating would need to explain this effect in a natural way.

The results described in this section are based on illustrative but
unrealistic models. The pressure differences introduced by the high temperature
regions would produce instabilities, and pressure equilibrium
would be reached in times given by the ratio of the width of the spikes to the speed of sound
in the gas ($\sim10$ km~s$^{-1}$). Hot regions with widths of $10^{15}$~cm, like
the ones in the model with thin spikes in Figure~\ref{fig3}, would have
dynamical lifetimes of $\sim30$ years. Heating and cooling processes have
similar timescales in ionized gas \citep{ost06}. Hence, the mechanism
responsible for the hot spikes should renew them within this timescale, as
previously noted by \citet{fer01}.\\

\section{Comments on models that include cold gas}

Models that reproduce the intensities of \ion{O}{2} RLs can also be built by
adding a component of cold gas that contributes significantly to the emission of
RLs but not to the emission of CELs.
This cold gas could arise from the photoionization of metal-rich inclusions
\citep[see][for \ion{H}{2} regions]{tsa05,sta07} or
could be gas ionized by X rays \citep{erc09} or cosmic rays \citep{gia05}.
Because the cold gas is poor in hydrogen or because its ionization fraction of
O$^{++}$ is larger than the one for H$^+$,
CELs will give better estimates of the metallicity in the models with cold gas
that have considered this issue \citep[see][]{sta07,erc09}.
To explain the ADF, the gas should have a significant fraction of O as O$^{++}$
and a temperature low enough to suppress the emission of CELs.
Some of the temperatures considered in the works mentioned above for the cold
component have values
$T_{\rm e}\ga4000$~K, not low enough to prevent some emission from CELs,
especially in the [\ion{N}{2}] lines
\citep[see, for example, Figure~2 in][]{tsa05}.
This emission will lower the value of
$T_{\rm e}$([\ion{N}{2}])/$T_{\rm e}$([\ion{O}{3}])
and hence can destroy the agreement between observations and models.
There are many uncertainties related to the physical conditions of the
hypothetical cold gas, and if its temperature is low enough, its emission will
not contribute to CELs significantly.
Hence, the observed values of $T_{\rm e}$([\ion{N}{2}])/$T_{\rm e}$([\ion{O}{3}])
can be used as a further constraint on this kind of explanations.\\

\section{Discussion}

Typical \ion{H}{2} regions have a zone close to the ionizing star where O$^{++}$
is the main ionization state of oxygen and a zone where O$^{+}$ dominates (and
where most of N$^{+}$ is also to be found).
The stellar radiation field that determines this ionization structure heats the
gas mainly through the ionization of neutral hydrogen.
The excess energy carried away by the ejected electrons increases with distance
from the ionizing star because the photoionization cross section of hydrogen is
strongly peaked at the threshold energy of 13.6~eV, and the preferred absorption
of photons with these energies hardens the ionizing radiation field.
The gas cools mainly through recombination and through the collisional excitation
of low-lying energy levels of abundant ions of heavy elements, responsible for
the production of CELs.
At each point, an equilibrium temperature is reached through the balance of all
the heating and cooling processes.
In this way, the temperature and the metallicity of the gas are inextricably
related.
The hardening of the radiation field at increasing distances from the ionizing
star and the very efficient cooling provided by the [\ion{O}{3}] lines produce
a temperature gradient that translates into the measured values
of $T_{\rm e}$([\ion{N}{2}])/$T_{\rm e}$([\ion{O}{3}])~$>1$ \citep{sta80,ost06}.
Finally, the total emission in CELs like [\ion{O}{2}] and [\ion{O}{3}] lines is
determined by both the ionization structure and the temperature structure of the
nebula.

The ADF found between the abundances implied by [\ion{O}{3}] CELs and \ion{O}{2}
RLs not only introduces
an uncertainty in the absolute value of the gas metallicity in observed nebulae;
it might also require for its explanation significant modifications of the
temperature structure predicted by simple photoionization models.
Therefore, a comparison between the temperature structures and metallicities
measured in observed objects and predicted by simple models can give indications
of how much modification the models can accommodate.

If we tried to build realistic models of the observed regions, we would need to
specify several, very difficult to know, pieces of information such as
the shape and luminosity of the ionizing
radiation field emitted by each massive star in the region, the distance of all
of these sources to the observed area, the amount and distribution of the
intervening material between the ionizing sources and the observed area, and the
density distribution of material in the observed area.
Instead, our approach has been to construct grids of simple models and to see
whether they reproduce the trends defined by complex real objects.
To minimize the uncertainties related to the measurement of line intensities,
we use some of the best spectra available for \ion{H}{2} regions. We compare
the temperatures, $T_{\rm e}$([\ion{N}{2}]) and $T_{\rm e}$([\ion{O}{3}]), and
the oxygen abundances derived with the observed CELs with those derived following
exactly the same procedure for grids of simple photoionization models with
parameters that should encompass the ones that characterize the observed objects.
The models are not tailored to reproduce any object; they just use the
best guesses in Cloudy for some parameters plus state-of-the-art model
atmospheres.
These models reproduce easily the observed values of the temperatures and the
temperature structure measured by the ratio
$T_{\rm e}$([\ion{N}{2}])/$T_{\rm e}$([\ion{O}{3}]) when their
input metallicities are close to those derived from CELs in the observed objects.

We also consider some modifications to the simple models so that they also
reproduce the observed \ion{O}{2} RLs.
We find that temperature fluctuations introduced by some extra heating
mechanism will destroy the agreement found for the simple models by decreasing
$T_{\rm e}$([\ion{N}{2}])/$T_{\rm e}$([\ion{O}{3}]) to values below the observed
ones.
This problem might be fixed if the extra heating mechanism happens to be more
efficient at larger distances from the ionizing star.
On the other hand, explanations of the ADF based on the addition of cold gas
in metal-rich inclusions or ionized by X rays or cosmic rays, will only work if
the temperature of this gas is so low that practically no CELs arise from
the inclusions.
None of these types of explanations of the ADF have been shown to work at the
moment, and without more details on the mechanism behind the discrepancy, we can
only provide these constraints that a successful explanation should meet.

In view of the very good agreement found between the simple models and the
observed objects, another explanation of the ADF in \ion{H}{2} regions should be
considered, namely, that the discrepancy is due to errors in the recombination
coefficients for \ion{O}{2}.
Some authors rule out this explanation because it would lead to
similar values of the ADF in all the objects, and though all \ion{H}{2} regions
and many planetary nebulae have $ADF(\mbox{O}^{++})\simeq2$,
some planetary nebulae have values that go up to 70 \citep{liu06}.
However, note that this does not rule out the possibility that the lowest values
of the ADF, found in \ion{H}{2} regions and many planetary nebulae, can be
explained in this way.
Consider the Orion nebula, M42, the \ion{H}{2} region with the best quality
spectrum and hence the one where the measured ADF is more reliable.
Multiplet 1 implies $ADF(\mbox{O}^{++})=1.5$ but other multiplets that should
give reliable values of the ADF imply $ADF(\mbox{O}^{++})=1.04\mbox{--}1.77$
\citep{est04}.
The same multiplets imply $ADF(\mbox{O}^{++})=1.5\mbox{--}2.3$ in NGC~3576 (in
the other nebulae only 1 to 3 reliable multiplets were measured).
If the recombination coefficients are underestimated by factors around or above
1.5 and if we also consider the uncertainties related to the measurement of these
very weak lines, the derived ADFs could be easily explained in \ion{H}{2} regions
and those planetary nebulae with similar ADFs.
A different explanation would only be needed for the handful of planetary
nebulae that show high ADFs.

On the other hand, the ADFs derived for other ions like C$^{++}$, although much
more uncertain than those derived for O$^{++}$, are usually larger than 1 and
correlate roughly with $ADF(\mbox{O}^{++})$ (see, e.g., Figure~18 in
\citealt{wan07} for results on planetary nebulae).
If the recombination coefficients for O$^{++}$ are underestimated, the same could
happen with these ions.
A revision of all these recombination coefficients would be valuable.

Real nebulae are likely to have heating mechanisms besides those considered in
models, and maybe some cold gas (due to shadows or ionized by alternative
sources).
This will introduce larger temperature fluctuations than those produced by simple
models.
In fact, the temperatures based on the Balmer or Paschen discontinuities
of hydrogen seem to be significantly lower than those derived using CELs in some
\ion{H}{2} regions \citep[see, e.g.,][]{gar04,gar06}.
However, in other \ion{H}{2} regions, the temperatures implied by the
hydrogen discontinuities and by CELs are fully consistent \citep{est04,gar05}.
The issue is far from being settled because the hydrogen temperatures are
difficult to measure and have large uncertainties.
But even if real objects have more temperature fluctuations than those predicted
by models, most of the ADF could still arise from errors in the recombination
coefficients.\\

\section{Conclusions}

We have constructed sets of simple photoionization models with parameters
that encompass those of nine \ion{H}{2} regions, chosen because
their observed spectra are of high quality.
Models and observations are analyzed in the same way to derive temperatures and
oxygen abundances using the relative intensities of CELs and hydrogen RLs.
The good agreement found between observations and models suggests that the models
capture the main characteristics of the observed objects.
In particular, the agreement between the derived and predicted values of
$T_{\rm e}$([\ion{N}{2}])/$T_{\rm e}$([\ion{O}{3}]) implies that the temperature
structure of the observed objects is reproduced by the models.
Since the [\ion{N}{2}] and [\ion{O}{3}] diagnostic lines used to derive
$T_{\rm e}$ will react differently to changes in the temperature structure,
this agreement places strong constraints on the modifications or new ingredients
that can be added to the models to explain the  \ion{O}{2} RLs.

The most straightforward explanation of the ADF in \ion{H}{2} regions (and those
planetary nebulae with $ADF(\mbox{O}^{++})\simeq2$) requires errors by this
amount in the recombination coefficients of \ion{O}{2} RLs.
However, we cannot rule out neither that there are relatively large temperature
fluctuations due to some unknown heating mechanism whose efficiency increases
with depth into the ionized cloud nor the presence of cold ionized gas if its
temperature is low enough to suppress all emission in CELs.
But whatever mechanism is invoked to explain the ADF, not only should it have a
plausible origin, it should also satisfy the stringent constraints imposed
by the observed temperature structures.\\

\acknowledgments
We thank C.~Morisset, R.~H.~Rubin, S.~Sim\'on-D\'\i az, G.~Stasi\'nska, and
G.~Tenorio-Tagle for useful comments and suggestions.
Bob Rubin also provided orange and violet.
We thank an anonymous referee for comments that helped to improve the paper.
This research has made use of NASA's Astrophysics Data System Bibliographic
Services.
The work was supported by Mexican CONACYT project 50359-F.\\

\end{document}